\documentclass[%
print,
 amsmath,amssymb,
 aps,
]{revtex4}

\usepackage[sc]{mathpazo}%此指令是令文章字体选用Palation.
\usepackage{graphicx}% Include figure files
\usepackage{dcolumn}% Align table columns on decimal point
\usepackage{bm}% bold math
\usepackage{dsfont}
\usepackage[landscape,papersize={297.1mm,210mm},left=1.9cm,right=1.6cm,top=2.7cm,bottom=2.8cm]{geometry}% 此指令是规定A4纸的页边距
\usepackage[                   %dvipdfm, pdflatex,pdftex这里决定运行文件的方式不同
            pdfstartview=FitH,
            colorlinks, %注释掉此项则交叉引用为彩色边框(将colorlinks和pdfborder同时注释掉)
            pdfborder=001,   %注释掉此项则交叉引用为彩色边框
            linkcolor=blue,
            anchorcolor=blue,
            citecolor=blue,
            urlcolor=blue
            ]{hyperref}
\usepackage{amsthm}
\makeatletter

\newcommand{\Rmnum}[1]{\expandafter\@slowromancap\romannumeral #1@}
\makeatother

\begin{document}
\title{Deterministic transformations of multilevel coherent states under coherence-preserving operations}
\author{Limei Zhang$^{1,2}$}
\author{Ting Gao$^1$}
\email{gaoting@hebtu.edu.cn}
\author{Fengli Yan$^3$}
\email{flyan@hebtu.edu.cn}
\affiliation{$^1$ School of Mathematical Sciences,
 Hebei Normal University, Shijiazhuang 050024, China \\
$^2$ Department of Mathematics and Computer Science, Hengshui Unversity, Hengshui 053000, China \\
$^3$ College of Physics, Hebei Key Laboratory of Photophysics Research and Application, Hebei Normal University, Shijiazhuang 050024, China}

\begin{abstract}
Quantum coherence, emerging from the 'superposition' of quantum states, is widely used in various information processing tasks. Recently, the resource theory of multilevel quantum coherence  is attracting substantial attention. In this paper, we mainly study the deterministic transformations of resource pure states via free operations in the theoretical framework for multilevel coherence. We prove that any two multilevel coherent resource pure states can be interconverted  with a nonzero probability via a completely positive and trace non-increasing $k$-coherence-preserving map. Meanwhile, we present the condition of the interconversions of two multilevel coherent resource pure states under  $k$-coherence-preserving operations.
In addition, we obtain that in the resource-theoretic framework of multilevel coherence, no resource state is isolated, that is, given a  multilevel coherent pure state $|\psi\rangle$, there exists another  multilevel coherent pure state $|\phi\rangle$ and a $k$-coherence-preserving operation $\Lambda_k$, such that $\Lambda_k(|\phi\rangle)=|\psi\rangle$.\\

\end{abstract}


\maketitle

\section{Introduction}
 Quantum coherence \cite{PRL113.140401,RMP89.041003}, one embodiment of the superposition principle of quantum states,  plays a central role in fundamental physics. It is widely used in various information processing tasks, such as  quantum cryptography \cite{RMP74.145}, metrology \cite{NP5.222,JPA47.424006}, thermodynamics \cite{NC6.6383,NC6.7689}, and even quantum biology \cite{JPCS302.012037}.
  Recently, the resource theory of quantum coherence  is flourishing \cite{PR762,PRL116.120404,RMP89.041003,PRL123.110402}. Within the standard
  framework of the resource theory of quantum coherence \cite{PRL113.140401}, the diagonal states in the prefixed reference basis are incoherent, and incoherent operations  are the free operations. While in the resource theory of quantum coherence, incoherent operation  is not the single and physically motivated choice of free operations that best describe the allowed state manipulations, unlike the standard choice of local operations and classical communication \big(LOCC\big) for the theory of entanglement \cite{QIC.7.1,RMP81.865,PR474}. Hence there has been extensive work on the operational properties and applications of quantum coherence under different sets of free operations \cite{PRL113.140401,PRL117.030401,PRL116.120404,PRX6.041028,PRA93.032326,Aberg2006}.

Quantum coherence is now recognized as a relatively  fledged resource \cite{RMP89.041003,PR762}, and great progress has been made on it. However, the coarse grained description in the most current literature is ultimately insufficient to reach fully understanding the essential role of the quantum superposition in the aforementioned  tasks. Thus, one needs to further consider the number of classical states in the coherent superposition, which leads to the concept of multilevel coherence \cite{NJP16.033007,PS90.074024,PRL116.080402}. The study of the rich structure of multilevel coherence has a tangible impact on many fields of physics \cite{NC6.6383,NJP16.033007,Nature543.647}. Ringbauer et al. \cite{PRX8.041007} developed the theoretical and experimental groundwork for characterizing and quantifying multilevel coherence, and present the robustness of multilevel coherence as a bona fide measure. Their results contribute to a better understanding of multilevel coherence. The theoretical framework \cite{PRX8.041007} consists of two basic ingredients: the set of multilevel coherence-free states which are defined by the coherence rank of the states, and the set of multilevel coherence-free operations.  The $k$-coherence-preserving operations \big($k\in\{1,2,\cdots,d\}$ in a $d$-dimensional quantum system\big),  which parallel to  the nonentangling operations \cite{NP.4.873(2008)} in entanglement theory,  belong to a class of  multilevel coherence-free operations \cite{PRX8.041007}.

In quantum resource theories, one of the most significant aspects is the rules governing state transformations via the free operations \cite{RMP89.041003,RMP81.865,RMP91.025001}.
For the state-to-state transformations, quantum coherence has similar features with quantum entanglement \cite{RMP81.865}.  The celebrated Nielsen theorem gave the necessary  and sufficient conditions for a class of entanglement transformations by LOCC \cite{PRL83.436}.  This  connects quantum entanglement to the linear-algebraic theory of majorization. Vidal \cite{PRL83.1046} generalized Nielsen's theorem and presented an optimal local conversion strategy of bipartite entangled pure states. Du et.al built the counterpart of Nielsen theorem for quantum coherence \cite{PRA91.052120} and also used the majorization as a key ingredient in the context of  the interconvertibility of coherent states.
The strategy \cite{PRL83.1046} was adapted  to the optimal conversion of coherent states under incoherent operations \cite{QIC15.1307}.
However, the situations change drastically for multipartite entanglement,  almost all pure multipartite entangled states are isolated under LOCC \cite{PRA93.012339,PRX8.031020}. To solve this problem,  the researchers relaxed the class of LOCC to wider free operations \cite{PRL122.120503}.
Accordingly, the interconversions of pure coherent states under different free operations were intensively investigated \cite{PRL117.030401,PRL116.120404,QIC15.1307,PRA94.052336,ShiSR2017,PRA97.052331}. Although many valuable results have been obtained, the deterministic  transformations of multilevel coherent states have remained relatively unexplored. We will focus on the deterministic  transformations of multilevel coherent states via $k$-coherence-preserving operations.

The paper is organized as follows.  We first briefly review the frameworks of coherence and multilevel coherence, and introduce two multilevel coherent measures: the robustness of multilevel coherence and the geometry measure of multilevel coherence in Sec.~\uppercase\expandafter{\romannumeral 2}. Then in Sec.~\uppercase\expandafter{\romannumeral 3}, we present our main result: that any two multilevel coherent resource pure states can be interconverted  with a nonzero probability via a completely positive and trace non-increasing $k$-coherence-preserving map, where the nonzero probability can be validly evaluated. Additionally, we answer the question given two resource pure states $|\psi_1\rangle$ and  $|\psi_2\rangle$, whether $|\psi_1\rangle$  can be transformed into $|\psi_2\rangle$ under  $k$-coherence-preserving operations, and also we prove no resource state is isolated.
Finally, we end with conclusions in Sec.~\uppercase\expandafter{\romannumeral 4}.

\section{Preliminaries}
Before discussing the deterministic  transformation of multilevel coherent states, it is instructive to review the general framework of the resource
theory of quantum coherence introduced in \cite{PRL113.140401}, and the framework of multilevel coherence \cite{PRX8.041007}. Throughout the paper, we consider the general $d$-dimensional Hilbert space $\mathcal{H}$. Let $\mathcal{D}(\mathcal{H})$ be the set of all density matrices on $\mathcal{H}$. Since coherence is  a basis dependent concept, we hereafter fix a particular basis, $\{|i\rangle\}_{i=1,\ldots,d}$, of $\mathcal{H}$.

Let $\mathcal{I}\subset\mathcal{D}(\mathcal{H})$ be the set of incoherent states. All incoherent density matrices $\delta\in\mathcal{I}$ are of the form
\begin{equation}
\begin{aligned}
\delta=\sum_{i=1}^{d}p_{i}|i\rangle\langle i|,
\end{aligned}
\end{equation}
where $p_i\in[0,1]$ and $\sum_ip_i=1$.

In the resource theory of quantum coherence, the definition of free operations is not unique. Within the framework of Baumgratz et al. \cite{PRL113.140401},  free operations are those of the incoherent operations, which act as a Kraus decomposition, i.e.,  $\Lambda(\cdot)=\sum\limits_n K_n\cdot K_n^{\dagger}$ and  the Kraus operators $\{K_n\}$  satisfy
$\sum\limits_n K_nK_n^{\dagger}=\mathds{I}_d$ and $K_n\mathcal{I}K_n^{\dagger}\subset\mathcal{I}$ for all $n$.

Another free operation in coherence theory is the incoherence-preserving operation \cite{PRA94.052324}. A quantum operation $\Lambda$ \big(a completely positive and trace-nonincreasing map\big) which maps incoherent quantum states  on the input space $\mathcal{I}_{in}$ to incoherent quantum states on the output space $\mathcal{I}_{out}$, is defined as incoherence-preserving operation. More succinctly,
\begin{equation}
 \rho\in\mathcal{I}_{in}\Rightarrow\Lambda(\rho)\in\mathcal{I}_{out}.
\end{equation}
Incoherence-preserving operations are the largest class of free operations for the resource
theory of quantum coherence and are parallel to  the nonentangling operations \cite{NP.4.873(2008)} in entanglement theory.

Equipped with the incoherent states and incoherent operations, Baumgratz et al. \cite{PRL113.140401} introduced the coherence measurement $\mathcal{C}$ mapping a quantum state $\rho$ to a nonnegative real number and satisfying

(C1) Nonnegativity, $\mathcal{C}(\rho)\geq 0$, and $\mathcal{C}(\delta)=0$ iff $\delta\in\mathcal{I}$.

(C2a) Monotonicity under incoherent operations, i.e., $\mathcal{C}(\Lambda[\rho])\leqslant \mathcal{C}(\rho)$ for any incoherent operation $\Lambda$.

(C2b) Monotonicity under  selective  incoherent operations on average, i.e.,
$\sum\limits_{i}p_{i}\mathcal{C}(\rho_{i})\leqslant \mathcal{C}(\rho)$ with probabilities $p_{i}=\text{Tr}[K_{i}\rho K_{i}^\dag]$, postmeasurement states $\rho_{i}=K_{i}\rho K_{i}^\dag/p_{i}$, and Kraus operators $\{K_{i}\}$.

(C3) Convexity, i.e., $\mathcal{C}(\sum\limits_{i}p_{i}\rho_{i})\leqslant \sum\limits_{i}p_{i}\mathcal{C}(\rho_{i})$
for any set of states $\{\rho_i\}$ and  probability distribution $\{p_i\}$.

Remarkably, conditions (C2b) and (C3) automatically imply condition (C2a) \cite{PRL113.140401}.

Similar to the  resource-theoretic framework of coherence, Ringbauer et al. \cite{PRX8.041007} developed the resource theory of multilevel coherence. In particular, they provided a new characterization of multilevel coherence-free states, free operations, rigorously unfolding the hierarchy of multilevel coherence. They also formalized the robustness of multilevel coherence \big(an efficiently computable measure of multilevel coherence\big).

\emph{Multilevel coherence-free quantum states}~~For a $d$-dimensional Hilbert space $\mathcal{H}$ and the aforementioned particular basis, $\{|i\rangle\}_{i=1,\ldots,d}$, any pure state $|\psi\rangle\in\mathcal{H}$ can be written as $|\psi\rangle=\sum\limits_{i=1}^d c_i|i\rangle$, with $\sum\limits_{i=1}^d |c_i|^2=1$. If there exist exactly $k$ nonzero coefficients $c_i$, then the coherence rank \cite{PRA96.042336} $r_c$ of $|\psi\rangle$ is $k$, i.e., $r_c(|\psi\rangle)= k$.
The sets $\mathcal{I}_k\subseteq\mathcal{D}(\mathcal{H})$ with $k\in\{1,2,\cdots,d\}$ are defined as all probabilistic mixtures of pure density operators $|\psi_i\rangle\langle\psi_i|$ with a coherence rank of at most $k$ \cite{PRA98.022328,PRX8.041007},
\begin{equation}
\mathcal{I}_k\equiv\left\{\sum\limits_ip_i|\psi_i\rangle\langle\psi_i|: p_i\geq0, \sum\limits_ip_i=1, r_c(|\psi_i\rangle)\leq k\right\}.
\end{equation}
Here $\mathcal{I}_k$ is the set of ($k$+1)-level coherence-free states. Note that,  the intermediate sets obey a strict hierarchy $\mathcal{I}_1\subset\mathcal{I}_2\subset\cdots\subset\mathcal{I}_d$ and are the free states in
the resource theory of multilevel coherence, where $\mathcal{I}_1$ is the set $\mathcal{I}$ of (fully) incoherent states and $\mathcal{I}_d\equiv\mathcal{D}(\mathcal{H})$ is the set of all states.  The coherence rank $r_c$, the number of nonzero coefficients $c_i$, reveals the multilevel nature of coherence.

\emph{Multilevel coherence-free operations}   In the resource theory of multilevel  coherence, the second ingredient is the set of free operations which are quantum operations that cannot create multilevel coherence. Generalizing the formalism introduced for standard coherence \cite{PRL113.140401,RMP89.041003}, one can refer to a linear completely positive and trace-preserving (CPTP) map $\Lambda$   as a $k$-coherence-preserving operation \cite{PRX8.041007} if it cannot  increase the coherence level, i.e., $\Lambda(\mathcal{I}_k)\subseteq\mathcal{I}_k$.

\emph{$k$-coherence-preserving map}~~A linear map $\Lambda$ is defined as a $k$-coherence-preserving map, if $\Lambda(\rho)/\mathrm{Tr}[\Lambda(\rho)]\in\mathcal{I}_k$, for any $\rho\in\mathcal{I}_k$.

\emph{Measure of multilevel coherence} In the resource theory of multilevel coherence, quantifying multilevel coherence is a crucial task. Here, we mainly introduce two well-defined measures of multilevel coherence: the robustness, and the geometric measure.

For a quantum state $\rho\in\mathcal{D}(\mathcal{H})$, the  robustness of $k$-coherence is defined as \cite{PRA98.022328}
 \begin{equation}\label{robustness}
R_k(\rho)=\min_{\delta\in \mathcal{I}_k}\left\{s\geq0~\Big|~\frac{\rho+s\delta}{1+s}\in\mathcal{I}_k\right\},
 \end{equation}
for $k\in\{2,3,\cdots,d\}$.

For a pure state $|\psi\rangle$, the geometric measure of $k$-coherence  is \cite{NJP20.033012}
\begin{equation}\label{geometric measure}
  G_k(|\psi\rangle)=1-\max_{|\phi\rangle\in\mathcal{I}_{k-1}}|\langle\phi|\psi\rangle|^2.
\end{equation}
Geometric measure of $k$-coherence is a computable quantifier of multilevel coherence, extending previous work \cite{NJP20.033012}.
\section{Main Results}
Multilevel quantum coherence  is a powerful, yet experimentally accessible quantum resource \cite{PRX8.041007}, and it is a key ingredient for practical applications from the transfer phenomena in many-body and complex systems to  quantum technologies \cite{PRX8.041007}. In the resource theory of multilevel quantum coherence, the study of transformations  of resource states via free operations is an important task. In this section, we mainly consider the transformation of multilevel coherent pure states. Given two multilevel coherent pure states, we study whether they can be interconverted via coherence-preserving maps/operations.

Let's start the main text with two very useful lemmas.

\emph{Lemma 1.} \cite{arxiv1711}~Suppose $\rho_1$, $\rho_2$ are density matrices, then there exists an operator $\Lambda$,
\begin{equation}
 \Lambda(\sigma)=\texttt{Tr}(A\sigma)\rho_1+\texttt{Tr}[(\mathbb{I}-A)\sigma]~\rho_2
\end{equation}
which is a CPTP map if $0\leq A\leq\mathbb{I}$. Here $\sigma\in \mathcal{D}(\mathcal{H})$.

\emph{Lemma 2.} For any quantum state $\sigma\in\mathcal{I}_k$ and multilevel coherent  pure states $\psi_1,~\psi_2\notin \mathcal{I}_k$ \big( $\psi_1$ and $\psi_2$ are the corresponding density matrices of $|\psi_1\rangle$ and $|\psi_2\rangle$, respectively\big), suppose $0< p\leq1$ and $\frac{1}{p}\left(\frac{1}{\texttt{Tr}(\psi_1\sigma)}-1\right)\geq R_{k}(\psi_2)$, then the map $\Lambda_k$
\begin{equation}\label{tiao jian}
  \Lambda_k(\sigma)=p~\texttt{Tr}(\psi_1\sigma)\psi_2+\texttt{Tr}[(\mathbb{I}-\psi_1)\sigma]~\delta,
\end{equation}
is a trace non-increasing $k$-coherence-preserving map. Here, $\delta\in\mathcal{I}_k$ is the optimal state achieving the robustness of $k$-coherence of $\psi_2$.
Specially, if $p=1$, then $\Lambda_k$ in Eq. (\ref{tiao jian}) is a $k$-coherence-preserving operation.

\emph{Proof.} For any quantum state $\sigma\in\mathcal{I}_k$, there is
\begin{equation}
\begin{array}{rl}
 \texttt{Tr}[\Lambda_k(\sigma)]&=p\texttt{Tr}(\psi_1\sigma)\texttt{Tr}(\psi_2)+\texttt{Tr}[(\mathbb{I}-\psi_1)\sigma]\texttt{Tr}(\delta)\\\\
&=\texttt{Tr}(\sigma)-(1-p)\texttt{Tr}(\psi_1\sigma)\\\\
&\leq \texttt{Tr}(\sigma),
\end{array} \end{equation}
with equality for $p=1$. Hence $\Lambda_k$ is a trace non-increasing map, while by \emph{Lemma} 1 it is a CPTP map for $p=1$.

Due to Eq. (\ref{tiao jian}), we have
\begin{equation}
     \Lambda_k(\sigma)\propto \psi_2+\frac{1}{p}\left(\frac{1}{\texttt{Tr}(\psi_1\sigma)}-1\right)\delta.
 \end{equation}
 If $\frac{1}{p}\left(\frac{1}{\texttt{Tr}(\psi_1\sigma)}-1\right)\geq R_k(\psi_2)$, then
\begin{equation}
     \Lambda_k(\sigma)\propto \psi_2+R_k(\psi_2)\delta+\left[\frac{1}{p}\left(\frac{1}{\texttt{Tr}(\psi_1\sigma)}-1\right)-R_k(\psi_2)\right]\delta,
 \end{equation}
where $R_k(\psi_2)$ denotes the robustness of $k$-coherence of $\psi_2$. Then according to the definition, one has
\begin{equation}
\frac{\psi_2+R_k(\psi_2)\delta}{1+R_k(\psi_2)}\in\mathcal{I}_k
 \end{equation}
with $\delta\in\mathcal{I}_k$,  thus $\Lambda_k(\sigma)/\texttt{Tr}[\Lambda_k(\sigma)]\in\mathcal{I}_k$.

So $\Lambda_k$ is a trace non-increasing and $k$-coherence-preserving map, while $\Lambda_k$ is a $k$-coherence-preserving operation for $p=1$. \qed

\emph{Theorem 1.} In the resource theory of multilevel quantum coherence, suppose free operations are $k$-coherence-preserving operations, and the resource states are quantum states not belonging to $\mathcal{I}_k$. All resource pure states are interconvertible with a non-zero probability. That is, for multilevel coherent pure states $\psi_1,~\psi_2\notin\mathcal{I}_k$, there exists a completely positive and trace non-increasing $k$-coherence-preserving map $\Lambda_k$ such that $\Lambda_k(\psi_1)=p\psi_2$ with $p\leq\frac{G_{k+1}(\psi_1)}{{R_k}(\psi_2)[1-G_{k+1}(\psi_1)]}$, $0<p\leq1$.

\emph{Proof.} For any  pure states $\psi_1,~\psi_2\notin \mathcal{I}_k$, let
\begin{equation}
  \Lambda_k(\cdot)=p~\texttt{Tr}(\psi_1\cdot)\psi_2+\texttt{Tr}[(\mathbb{I}-\psi_1)\cdot]~\delta,
\end{equation}
where $\delta$ is the optimal state achieving the robustness of $\psi_2$.
By \emph{Lemma} 1 and the proof of \emph{Lemma} 2, $\Lambda_k$ is a completely positive and trace non-increasing map.

Next we  prove that $\Lambda_k$ is a $k$-coherence-preserving map.
From the definition of geometric measure, for a multilevel coherent pure state $\psi_1\notin \mathcal{I}_k$ and any $\sigma\in\mathcal{I}_k$, one has
\begin{equation}\label{geometric inequality}
\begin{array}{rl}
1-G_{k+1}(\psi_1)&=\max\limits_{|\phi\rangle\in\mathcal{I}_k}|\langle\phi|\psi_1\rangle|^2\\\\
&=\max\limits_{|\phi\rangle\in\mathcal{I}_k}\texttt{Tr}(\psi_1|\phi\rangle\langle\phi|)\\\\
&\geq \texttt{Tr}(\psi_1\sigma).
\end{array} \end{equation}

Note that $G_{k+1}(\psi_1)<1$, if we can choose  $p\leq\frac{G_{k+1}(\psi_1)}{R_k(\psi_2)[1-G_{k+1}(\psi_1)]}$, and $0< p\leq1$, then
\begin{equation}
\begin{array}{rl}
R_k(\psi_2)&\leq\frac{G_{k+1}(\psi_1)}{p[1-G_{k+1}(\psi_1)]}\\\\
&=\frac{1}{p}\left(\frac{1}{1-G_{k+1}(\psi_1)}-1\right)\\\\
&\leq\frac{1}{p}\left(\frac{1}{\texttt{Tr}(\psi_1\sigma)}-1\right),
\end{array} \end{equation}
where the last inequality is due to  inequality (\ref{geometric inequality}). Thus, it follows from \emph{Lemma} 2 that $\Lambda_k$  is a  $k$-coherence-preserving map,  and
\begin{equation}
  \Lambda_k(\psi_1)=p~\texttt{Tr}(\psi_1\psi_1)\psi_2+\texttt{Tr}\big[(\mathbb{I}-\psi_1)\psi_1\big]~\delta=p\psi_2.
\end{equation}
This concludes the proof. \qed

For a pure state $|\psi\rangle=(\nu_1,\nu_2,\cdots,\nu_d)^T$ with real entries satisfying $\nu_1\geq\nu_2\geq\cdots\geq\nu_d\geq0$, an analytical formula for the robustness of $k$-coherence \big($k\in\{2,3,\cdots,d\}$\big)
\begin{equation}\label{robustness}
R_k(|\psi\rangle\langle\psi|)=\frac{s_l^2}{k-l+1}-\sum_{i=l}^d\nu_i^2,
 \end{equation}
was derived in \cite{PRA98.022328}, where $l\in\{2,3,\cdots,k\}$ is the largest integer such that $\nu_{l-1}\geq\frac{s_l}{k-l+1}$ with $s_l\equiv\sum\limits_{i=l}^d\nu_i$ \big(if no such integer exists, then set $l$=1\big).  If $|\psi\rangle$ is not of this form then it  can be converted to this form via a diagonal unitary and/or a permutation matrix, and these operations do not affect the value of  $R_k$.

Regula et al. \cite{NJP20.033012} obtained a closed formula for  the geometric measure of $k$-coherence of arbitrary pure states $|\phi\rangle$ as
\begin{equation}\label{geometric measure}
  G_k(|\phi\rangle)=1-\sum_{i=1}^{k-1}|\mu^{\downarrow}_i|^2,
\end{equation}
where $\mu^{\downarrow}_i$ is the $i$th largest coefficient (by absolute value) of $|\phi\rangle$.

Using Eqs. (\ref{robustness}) and (\ref{geometric measure}), the bound in \emph{Theorem 1} can be easily evaluated. Suppose  $|\psi_1\rangle=(\mu_1,\mu_2,\cdots,\mu_d)^T$ with $|\mu_1|\geq|\mu_2|\geq\cdots\geq|\mu_d|$ and $|\psi_2\rangle=(\nu_1,\nu_2,\cdots,\nu_d)^T$ with  real entries satisfying $\nu_1\geq\nu_2\geq\cdots\geq\nu_d\geq0$ in \emph{Theorem 1}, then one can validly evaluate $p$. That is, for multilevel coherent pure states $\psi_1,~\psi_2\notin \mathcal{I}_k$, there exists a completely positive and trace non-increasing $k$-coherence-preserving map $\Lambda_k$ such that $\Lambda_k(\psi_1)=p\psi_2$ with
 $p\leq\frac{\big(k-l+1\big)\left(1-\sum\limits_{i=1}^k|\mu_i|^2\right)}{\left[s_l^2-(k-l+1)(\sum\limits_{i=l}^n\nu_i^2)\right]\left(\sum\limits_{i=1}^k|\mu_i|^2\right)}$, where $l\in\{2,3,\cdots,k\}$ is the largest integer such that $\nu_{l-1}\geq\frac{s_l}{k-l+1}$ with $s_l\equiv\sum\limits_{i=l}^n\nu_i$, and $0< p\leq1$. This is useful as these bounds are often not easy to compute for other resources, e.g. in multipartite entanglement theory.

\emph{Theorem 2.}  For the resource theory of multilevel coherence, $k$-coherence-preserving operations are free operations.  Then no resource state is isolated. For any resource pure state $\psi\notin \mathcal{I}_k$, there always exists another resource pure state $\phi\notin \mathcal{I}_k$, and a $k$-coherence-preserving operation $\Lambda_k$ such that $\Lambda_k(\phi)=\psi$.

\emph{Proof.}  For the resource pure state $\psi\notin \mathcal{I}_k$, $R_k(\psi)$ is the robustness of $k$-coherence of $\psi$. If the geometric measure of $k+1$-coherence of the resource pure state $\phi$  satisfies
\begin{equation}
  \frac{G_{k+1}(\phi)}{R_k(\psi)[1-G_{k+1}(\phi)]}\geq 1,
\end{equation}
that is, $G_{k+1}(\phi)\geq1-\frac{1}{R_k(\psi)+1}$, then from \emph{Lemma} 2, one can construct a $k$-coherence-preserving operation
\begin{equation}
  \Lambda_k(\cdot)=\texttt{Tr}(\phi\cdot)\psi+\texttt{Tr}[(\mathbb{I}-\phi)\cdot]~\delta,
\end{equation}
where $\delta$ is the optimal state achieving the robustness of $k$-coherence of $\psi$. It implies that
\begin{equation}
  \Lambda_k(\phi)=\texttt{Tr}(\phi\phi)\psi+\texttt{Tr}[(\mathbb{I}-\phi)\phi]~\delta=\psi,
\end{equation}
which completes the proof. \qed

\emph{Theorem 3.}  In the resource theory of multilevel quantum coherence, $k$-coherence-preserving operations are the free operations. For any resource pure state $\psi_1, \psi_2\notin \mathcal{I}_k$, $\psi_1$ can be transformed to $\psi_2$ via  a $k$-coherence-preserving operation $\Lambda_k$ if  the geometric measure of $k+1$-coherence of $\psi_1$ and the robustness of $k$-coherence of $\psi_2$ satisfy the condition $\frac{G_{k+1}(\psi_1)}{R_k(\psi_2)[1-G_{k+1}(\psi_1)]}\geq 1$.

\emph{Proof.} Let \begin{equation}
  \Lambda_k(\cdot)=\texttt{Tr}(\psi_1\cdot)\psi_2+\texttt{Tr}[(\mathbb{I}-\psi_1)\cdot]~\delta,
\end{equation}
where $\delta$ is the optimal state achieving the robustness of $k$-coherence of $\psi_2$.

If the geometric measure of $k+1$-coherence of $\psi_1$ and the robustness of $k$-coherence of $\psi_2$ satisfy
\begin{equation}
  \frac{G_{k+1}(\psi_1)}{R_k(\psi_2)[1-G_{k+1}(\psi_1)]}\geq 1,
\end{equation}
then by \emph{Lemma} 2, $\Lambda_k$ is a $k$-coherence-preserving operation and
\begin{equation}
  \Lambda_k(\psi_1)=\texttt{Tr}(\psi_1\psi_1)\psi_2+\texttt{Tr}\big[(\mathbb{I}-\psi_1)\psi_1\big]~\delta=\psi_2.
\end{equation}
This proof is complete. \qed

\emph{Corollary.} In the theoretical framework for multilevel coherence, $k$-coherence-preserving operations are the free operations. The coherence state $\psi=|\psi\rangle\langle\psi|$ where  $|\psi\rangle=\frac{1}{\sqrt{d}}\sum\limits_{i=1}^d|i\rangle$ can be transformed to any other resource state $\phi\notin\mathcal{I}_k$ via $k$-coherence-preserving operations, for any $k$.

%\emph{Proof.} $G_k(\psi)=1-\frac{k}{d}$, and for any $\phi$, $R_k(\phi)\leq\frac{d}{k}-1$. Hence we can obtain $\frac{G_k(\psi)}{R_k(\phi)[1-G_k(\psi)]}\geq 1$. Due to \emph{Theorem 3}, $\psi$ can transform to any state $\phi$.

Remarkably, the proof of \emph{Theorem 1} actually gives the explicit construction of a completely positive and trace non-increasing $k$-coherence-preserving map  $\Lambda_k$ such that $\Lambda_k(\psi_1)=p\psi_2$ when the nonzero probability meets
$p\leq\frac{G_{k+1}(\psi_1)}{{R_k}(\psi_2)[1-G_{k+1}(\psi_1)]}$. If pure states $\psi_1,~\psi_2\notin\mathcal{I}_k$, we have $R_k(\psi_2)>0$ and $0<G_{k+1}(\psi_1)<1$, hence there always exists $p\in(0,1]$ such that  probability condition holds. The \emph{Theorems} 2 and 3, to some extent, are corollaries. In \emph{Theorems 2}, for a pure state $\psi\notin\mathcal{I}_k$, one can always find a pure state $\phi\notin\mathcal{I}_k$ with
$G_{k+1}(\phi)$ large enough to satisfy the required condition. Given pure states $\psi_1, \psi_2\notin \mathcal{I}_k$, if $R_k(\psi_2)$ is sufficiently smaller than $G_{k+1}(\psi_1)$, then one can pick $p=1$ in the \emph{Theorem 1}, such that $\psi_1$ is transformed to $\psi_2$ via a $k$-coherence-preserving operation.
\section{Conclusion}
The study of multilevel quantum coherence reveals further parallels between the resource theories of coherence and entanglement. The characterization  and quantification of multilevel coherence can apply semidefinite programming rather than general convex optimization. Hence it verifies that multilevel quantum coherence is a powerful, yet experimentally accessible quantum resource.
%The resource theory of multilevel quantum coherence has some significant operational advantages over the standard resource theory of quantum coherence.
In this paper, we focus on the transformation of multilevel coherent resource pure states which has remained relatively unexplored.
In the theoretical framework for multilevel coherence, the free operations are  $k$-coherence-preserving operations, we show that any two multilevel coherent resource pure states can be interconverted with a nonzero probability via a completely positive and trace non-increasing $k$-coherence-preserving map. Here the nonzero probability is related with the robustness and the geometric measure of multilevel coherence. We present the condition of the interconversions of any two multilevel coherent resource pure states under  $k$-coherence-preserving operations.  Moreover, we obtain that in the resource theory of multilevel quantum coherence, no resource state is isolated. Finally, we  show that in the resource theory of multilevel quantum coherence, the coherence state $|\psi\rangle=\frac{1}{\sqrt{d}}\sum\limits_{i=1}^d|i\rangle$ can be transformed to any other multilevel coherent pure state under $k$-coherence-preserving operations. We expect that our approach can be used to future study the framework of multilevel coherence and  interconversion of resource states.
\begin{acknowledgments}
This work was supported by the National Natural Science Foundation of China under Grant No.12071110,
the Hebei Natural Science Foundation of China under Grant Nos. A2020205014 and A2018205125, and the Education Department of Hebei Province Natural Science Foundation under Grant No. ZD2020167.
\end{acknowledgments}


\begin{thebibliography}{99}

%1-10
%Z. Ficek, S. S. Swain, Quantum interference and coherence: theory and experiments, Springer series in optical sciences, 2004.


\bibitem{PRL113.140401}T. Baumgratz, M. Cramer, and M. B. Plenio, Quantifying coherence, \href{https://journals.aps.org/prl/abstract/10.1103/PhysRevLett.113.140401} {Phys. Rev. Lett. \textbf{113}, 140401 (2014)}.
\bibitem{RMP89.041003}A. Streltsov, G. Adesso, and M. B. Plenio, Quantum coherence as a resource,  \href{https://journals.aps.org/rmp/abstract/10.1103/RevModPhys.89.041003} {Rev. Mod. Phys. \textbf{89}, 041003 (2017)}.

\bibitem{RMP74.145}N. Gisin, G. Ribordy, W. Tittel, and H. Zbinden,  Quantum cryptography,  \href{https://journals.aps.org/rmp/abstract/10.1103/RevModPhys.74.145} {Rev. Mod. Phys. \textbf{74}, 145 (2002)}.
  \bibitem{NP5.222}V. Giovannetti, S. Lloyd, and L. Maccone,  Advances in quantum metrology,  \href{https://www.nature.com/articles/nphoton.2011.35#citeas} { Nat. Photon. \textbf{5}, 222 (2011)}.
 \bibitem{JPA47.424006}G. T$\acute{o}$th and I. Apellaniz,  Quantum metrology from a quantum information science perspective,  \href{https://iopscience.iop.org/article/10.1088/1751-8113/47/42/424006/meta} {J. Phys.
A: Math. Theor. \textbf{47}, 424006 (2014)}.
\bibitem{NC6.6383}M. Lostaglio, D. Jennings, and T. Rudolph,  Description of quantum coherence in thermodynamic processes requires constraints beyond free energy,  \href{https://www.nature.com/articles/ncomms7383} { Nat. Commun. \textbf{6}, 6383 (2015)}.
\bibitem{NC6.7689}V. Narasimhachar and G. Gour,  Low-temperature thermodynamics with quantum coherence,  \href{https://www.nature.com/articles/ncomms8689} { Nat. Commun. \textbf{6}, 7689 (2015)}.
\bibitem{JPCS302.012037} S. Lloyd,  Quantum coherence in biologial systems,  \href{https://doi.org/10.1088/1742-6596/302/1/012037} { J. Phys.: Conf. Ser. \textbf{302}, 012037 (2011)}.
\bibitem{PRL116.120404}A. Winter and D. Yang, Operational resource theory of coherence, \href{https://journals.aps.org/prl/abstract/10.1103/PhysRevLett.116.120404} {Phys. Rev. Lett. \textbf{116}, 120404 (2016)}.
\bibitem{PR762}M. L. Hu, X. Y. Hu, J. C. Wang, Y. Peng, Y. R. Zhang, and H. Fan, Quantum coherence and geometric quantum discord,  \href{https://doi.org/10.1016/j.physrep.2018.07.004} {Phys. Rep. \textbf{762-764}, 1 (2018)}.
\bibitem{PRL123.110402}F. Bischof,  H. Kampermann, and D. Bru${\ss}$, Resource theory of coherence based on positive-operator-valued measures, \href{https://journals.aps.org/prl/abstract/10.1103/PhysRevLett.123.110402} {Phys. Rev. Lett. \textbf{123}, 110402 (2019)}.
\bibitem{QIC.7.1} M. B. Plenio and S. Virmani, An introduction to entanglement measures,  \href{https://dl.acm.org/doi/10.5555/2011706.2011707} {Quantum Inf. Comput. \textbf{7}, 1 (2007)}.
\bibitem{RMP81.865}R. Horodecki, P. Horodecki, M. Horodecki, and K. Horodecki, Quantum entanglement,  \href{https://journals.aps.org/rmp/abstract/10.1103/RevModPhys.81.865} {Rev. Mod. Phys. \textbf{81}, 865 (2009)}.
\bibitem{PR474}O. G\"{u}hne and G. T\'{o}th, Entanglement detection,  \href{https://www.sciencedirect.com/science/article/pii/S0370157309000623} {Phys. Rep. \textbf{474}, 1 (2009)}.
\bibitem{PRL117.030401}E. Chitambar and G. Gour, Critical examination of incoherent operations and a physically consistent resource theory of quantum coherence, \href{https://journals.aps.org/prl/abstract/10.1103/PhysRevLett.117.030401} {Phys. Rev. Lett. \textbf{117}, 030401 (2016)}.
\bibitem{PRX6.041028}B. Yadin, J. J. Ma, D. Girolami, M. Gu, and V. Vedral, Quantum processes which do not use coherence, \href{https://journals.aps.org/prx/abstract/10.1103/PhysRevX.6.041028} {Phys. Rev. X \textbf{6}, 041028 (2016)}.
\bibitem{PRA93.032326}Y. Peng, Y. Jiang, and H. Fan, Maximally coherent states and coherence-preserving operations, \href{https://journals.aps.org/pra/abstract/10.1103/PhysRevA.93.032326} {Phys. Rev. A \textbf{93}, 032326 (2016)}.
\bibitem{Aberg2006} J. ${\AA}$berg, Quantifying superposition, \href{https://arxiv.org/abs/quant-ph/0612146} {arXiv:quant-ph/0612146}.
\bibitem{NJP16.033007} F. Levi and F. Mintert, A quantitative theory of coherent delocalization, \href{https://iopscience.iop.org/article/10.1088/1367-2630/16/3/033007} {New J. Phys. \textbf {16}, 033007 (2014)}.
\bibitem{PS90.074024} J. Sperling and W. Vogel, Convex ordering and quantification of quantumness, { Phys. Scr. \textbf{90}, 074024 (2015)}.
\bibitem{PRL116.080402} N. Killoran, F. E. Steinhoff, and M. B. Plenio, Converting nonclassicality into entanglement, \href{https://journals.aps.org/prl/abstract/10.1103/PhysRevLett.116.080402} {Phys. Rev. Lett. \textbf{116}, 080402 (2016)}.
\bibitem{Nature543.647} G. D. Scholes, \textit{et al.}, Using coherence to enhance function in chemical and biophysical systems, \href{https://www.nature.com/articles/nature21425} { Nature (London) \textbf{543}, 647 (2017)}.%作者有19人%
\bibitem{PRX8.041007}M. Ringbauer, T. R. Bromley, M. Cianciaruso, L. Lami, W. Y. S. Lau, G. Adesso, A. G. White, A. Fedrizzi, and M. Piani, Certification and quantification of multilevel quantum coherence, \href{https://journals.aps.org/prx/abstract/10.1103/PhysRevX.8.041007} {Phys. Rev. X \textbf{8}, 041007 (2018)}.
\bibitem{NP.4.873(2008)}F. G. Brandao and M. B. Plenio, Entanglement theory and the second law of thermodynamics, \href{https://www.nature.com/articles/nphys1100} {Nat. Phys. \textbf{4}, 873 (2008)}.
\bibitem{RMP91.025001}E. Chitambar and G. Gour, Quantum resource theories,  \href{https://journals.aps.org/rmp/abstract/10.1103/RevModPhys.91.025001} {Rev. Mod. Phys. \textbf{91}, 025001 (2019)}.
\bibitem{PRL83.436}M. A. Nielsen, Conditions for a class of entanglement transformations, \href{https://journals.aps.org/prl/abstract/10.1103/PhysRevLett.83.436} {Phys. Rev. Lett. \textbf{83}, 436 (1999)}.
\bibitem{PRL83.1046}G. Vidal, Entanglement of pure states for a single copy, \href{https://journals.aps.org/prl/abstract/10.1103/PhysRevLett.83.1046} {Phys. Rev. Lett. \textbf{83}, 1046 (1999)}.
\bibitem{PRA91.052120}S. P. Du, Z. F. Bai, and Y. Guo, Conditions for coherence transformations under incoherent operations,  \href{https://journals.aps.org/pra/abstract/10.1103/PhysRevA.91.052120} {Phys. Rev. A \textbf{91}, 052120 (2015)}; Erratum \href{https://journals.aps.org/pra/abstract/10.1103/PhysRevA.95.029901} {Phys. Rev. A \textbf{95}, 029901 (2017)}.
\bibitem{QIC15.1307} S. P. Du, Z. F. Bai, and X. F. Qi, Coherence measures and optimal conversion for coherent states, \href{https://www.researchgate.net/publication/274967347_Coherence_measures_and_optimal_conversion_for_coherent_states} {Quantum Info. Comput. \textbf{15-16}, 1307 (2015)}.
\bibitem{PRA93.012339}M. Hebenstreit, C. Spee, and B. Kraus, Maximally entangled set of tripartite qutrit states and pure state separable transformations which are not possible via local operations and classical communication, \href{https://journals.aps.org/pra/abstract/10.1103/PhysRevA.93.012339} {Phys. Rev. A \textbf{93}, 012339 (2016)}.
\bibitem{PRX8.031020}D. Sauerwein, N. R. Wallach, G. Gour, and B. Kraus, Transformations among pure multipartite entangled states via local operations
are almost never possible, \href{https://journals.aps.org/prx/abstract/10.1103/PhysRevX.8.031020} {Phys. Rev. X \textbf{8}, 031020 (2018)}.
\bibitem{PRL122.120503}P. C. Tejada,  C. Palazuelos, and J. I. de Vicente, Resource theory of entanglement with a unique multipartite maximally entangled state, \href{https://journals.aps.org/prl/abstract/10.1103/PhysRevLett.122.120503} {Phys. Rev. Lett. \textbf{122}, 120503 (2019)}.
 \bibitem{PRA94.052336}E. Chitambar and G. Gour, Comparison of incoherent operations and measures of coherence, \href{https://journals.aps.org/pra/abstract/10.1103/PhysRevA.94.052336} {Phys. Rev. A \textbf{94}, 052336 (2016)}. Erratum: \href{https://journals.aps.org/pra/abstract/10.1103/PhysRevA.95.019902} {Phys. Rev. A \textbf{95}, 019902 (2017)}.
\bibitem{ShiSR2017}H. L. Shi, X. H. Wang, S. Y. Liu, W. L. Yang, Z. Y. Yang, and  H. Fan, Coherence transformations in
single qubit systems,  \href{https://www.nature.com/articles/s41598-017-13687-4} {Sci.
Rep. \textbf{7}, 14806 (2017)}.
\bibitem{PRA97.052331}G. Torun and A. Yildiz, Deterministic transformations of coherent states under incoherent operations, \href{https://journals.aps.org/pra/abstract/10.1103/PhysRevA.97.052331} {Phys. Rev. A \textbf{97}, 052331 (2018)}.
\bibitem{PRA94.052324}I. Marvian and R. W. Spekkens, How to quantify coherence: Distinguishing speakable and unspeakable notions, \href{https://journals.aps.org/pra/abstract/10.1103/PhysRevA.94.052324} {Phys. Rev. A \textbf{94}, 052324 (2016)}.
\bibitem{PRA96.042336}S. Chin, Coherence number as a discrete quantum resource, \href{https://journals.aps.org/pra/abstract/10.1103/PhysRevA.96.042336} {Phys. Rev. A \textbf{96}, 042336 (2017)}.
\bibitem{PRA98.022328}N. Johnston, C. K. Li, S. Plosker, Y. T. Poon, and B. Regula, Evaluating the robustness of $k$-coherence and $k$-entanglement, \href{https://journals.aps.org/pra/abstract/10.1103/PhysRevA.98.022328} {Phys. Rev. A \textbf{98}, 022328 (2018)}.
\bibitem{NJP20.033012} B. Regula, M. Piani, M. Cianciaruso, T. R. Bromley, A. Streltsov, and G, Adesso, Converting multilevel nonclassicality into genuine
multipartite entanglement, \href{https://iopscience.iop.org/article/10.1088/1367-2630/aaae9d} { New J. Phys. \textbf{20}, 033012 (2018)}.
\bibitem{arxiv1711}E. Chitambar, J. I. de Vicente, M. W. Girard, and G. Gour, Entanglement manipulation and distillability beyond LOCC, \href{http://cn.arxiv.org/abs/1711.03835v2} {arXiv:quant-ph/1711. 03835}.

\end{thebibliography}
\end{document}